\documentclass[lettersize,journal]{IEEEtran}
\usepackage{amsmath,amsfonts}
\usepackage{algorithmic}
\usepackage{array}
\usepackage{subcaption} 
\usepackage{textcomp}
\usepackage{stfloats}
\usepackage{url}
\usepackage{verbatim}
\usepackage{graphicx}
\def\BibTeX{{\rm B\kern-.05em{\sc i\kern-.025em b}\kern-.08em
    T\kern-.1667em\lower.7ex\hbox{E}\kern-.125emX}}
\usepackage{balance}
\usepackage{booktabs}
\usepackage{multirow}
\usepackage{caption}
\usepackage{float}
\usepackage{cite}
\usepackage{hyperref}
\usepackage{comment}
\usepackage{pgfplots} 
\pgfplotsset{compat=1.17} 
\usepackage{algorithmic}
\usepackage{wasysym}  
\usepackage{stfloats}
\usepackage{xcolor}
\usepackage{makecell}
\usepackage[linesnumbered,ruled,vlined]{algorithm2e} 
\usepackage[margin=1.5cm]{geometry}
\SetKwInOut{Parameter}{Parameter}
\hypersetup{
    colorlinks=true,
    linkcolor=blue,
    filecolor=magenta,      
    urlcolor=cyan,
}
\usepackage{hyperref}
\usepackage[multiple]{footmisc}
\usepackage{amsmath}
\usepackage{tikz}
\usetikzlibrary{shapes}

\usepackage[multiple]{footmisc}

\begin{document}
\title{IDMap: A Pseudo-Speaker Generator Framework Based on Speaker Identity Index to Vector Mapping}

\author{Zeyan Liu, Liping Chen,  \IEEEmembership{Senior Member, IEEE}, Kong Aik Lee,  \IEEEmembership{Senior Member, IEEE}, \\ and Zhenhua Ling \IEEEmembership{Senior Member, IEEE}
\thanks{Zeyan Liu, Liping Chen, and Zhenhua, Ling are with the University of Science and Technology of China, China (e-mail: xy671231@mail.ustc.edu.cn, lipchen@ustc.edu.cn, zhling@ustc.edu.cn). Kong Aik Lee is with the Hong Kong Polytechnic University, China (e-mail: kong-aik.lee@polyu.edu.hk). }
\thanks{\emph{Corresponding author: Liping Chen.}}
\thanks{This work was supported in part by the National Natural Science Foundation of China under Grant U23B2053, and the Fundamental Research Funds for the Central Universities WK2100000043.}}

\markboth{Journal of \LaTeX\ Class Files,~Vol.~18, No.~9, September~2020}%
{How to Use the IEEEtran \LaTeX \ Templates}

\maketitle

\begin{abstract}

Facilitated by the speech generation framework that disentangles speech into content, speaker, and prosody, voice anonymization is accomplished by substituting the original speaker embedding vector with that of a pseudo-speaker. In this framework, the pseudo-speaker generation forms a fundamental challenge. Current pseudo-speaker generation methods demonstrate limitations in the uniqueness of pseudo-speakers, consequently restricting their effectiveness in voice privacy protection. Besides, existing model-based methods suffer from heavy computation costs. Especially, in the large-scale scenario where a huge number of pseudo-speakers are generated, the limitations of uniqueness and computational inefficiency become more significant. To this end, this paper proposes a framework for pseudo-speaker generation, which establishes a mapping from speaker identity index to speaker vector in the feedforward architecture, termed IDMap. Specifically, the framework is specified into two models: IDMap-MLP and IDMap-Diff. Experiments were conducted on both small- and large-scale evaluation datasets. Small-scale evaluations on the LibriSpeech dataset validated the effectiveness of the proposed IDMap framework in enhancing the uniqueness of pseudo-speakers, thereby improving voice privacy protection, while at a reduced computational cost. Large-scale evaluations on the MLS and Common Voice datasets further justified the superiority of the IDMap framework regarding the stability of the voice privacy protection capability as the number of pseudo-speakers increased. Audio samples and open-source code can be found in \url{https://github.com/VoicePrivacy/IDMap}.
\end{abstract}


\renewcommand{\arraystretch}{1.6} 

\section{Introduction}

\IEEEPARstart{I}{n} recent years, with the significant advancements in speech technologies \cite{li2022recent, ning2019review, casanova2022yourtts, voice_conversion_overview, snyder17_interspeech, ECAPATDNN}, the potential misuse of information conveyed Aby speech has led to a rise in security threats. Particularly, the malicious exploitation of speaker attributes leads to violations of voice privacy, calling for the development of voice privacy protection techniques. Among them, the voice anonymization technique, driven by the speech generation framework based on attributes disentanglement, offers a viable solution by replacing the original speaker attributes with those of a pseudo-speaker \cite{xvector-based}. In this technique, the speaker attribute within the original speech is disentangled and represented with a speaker embedding vector, e.g., the x-vector\cite{snyder17_interspeech}. The pseudo-speaker vector is subsequently derived and used to replace the original speaker vector, facilitating the generation of anonymized speech with the pseudo-speaker vector.

In this technique, the generation of pseudo-speakers poses a fundamental challenge. Currently, pseudo-speaker generation methods can be classified into three categories: reference pool-based methods \cite{nac,vq-bn,cohort-speaker,psd}, transformation-based methods \cite{ohnn,svd}, and generative methods \cite{turner2022generating,GAN}. In the reference pool-based methods, a reference pool is predefined. A pseudo-speaker may be randomly selected from the reference pool \cite{nac,vq-bn}. Besides, a cohort speaker set can be chosen from the reference pool to derive the pseudo-speaker. For instance, the pseudo-speaker vector is derived by averaging the vectors of cohort speakers \cite{average}. A pseudo-speaker distribution (PSD) estimator is trained from the utterances of the cohort speakers, with the pseudo-speaker represented by the resultant distribution \cite{psd}. In the transformation-based methods, the speaker embedding vector extracted from the original utterance is processed and transformed into an anonymized version as the pseudo-speaker vector. To name a few, the learnable orthogonal Householder (LOH) method \cite{ohnn} applies a neural network-parameterized orthogonal transformation to rotate an original speaker vector to get the corresponding pseudo-speaker \cite{ohnn}. The singular value decomposition (SVD)-based method \cite{svd} decomposes the original speaker vector using the SVD algorithm and subsequently applies iterative non-linear transformations to the singular values, thereby generating the speaker vector as the pseudo-speaker vector. In the generative methods, the pseudo-speaker vector is generated via sampling from a Gaussian mixture distribution \cite{turner2022generating} and with a generative adversarial network (GAN) \cite{GAN}.


Usually, the requirements on the pseudo-speakers include de-identification and uniqueness. The de-identification attribute necessitates that the pseudo-speaker differs from the original, while uniqueness requires that a specific pseudo-speaker is distinct from others. To date, existing pseudo-speaker generation methods have demonstrated satisfactory de-identification capabilities. This paper focuses on the uniqueness attribute of the pseudo-speaker, along with the computation efficiency of its generation process. 
In the uniqueness dimension, the reference pool-based, the generative methods, and the SVD-based method in the transformation-based methods lack constraints on the distinctiveness among the pseudo-speakers during their generation, thereby limiting their uniqueness. Especially when the anonymization is conducted at the utterance level, different utterances from the same original speaker may be assigned the same pseudo-speaker or pseudo-speakers with high similarity, rendering the anonymized utterances linkable in speaker identity. The LOH method achieves uniqueness through the loss function definition. However, it is limited to speaker-level anonymization, rendering it incapable of assigning a unique pseudo-speaker to every single original utterance. In terms of computational efficiency, the model-based pseudo-speaker generation methods, wherein models are developed to generate pseudo-speaker vectors, incur significant computational costs. Specifically, the PSD method trains a network-based distribution estimator for each pseudo-speaker, making it quite time-consuming. Besides, the transformation-based and generative methods operate iteratively, causing a heavy computation cost. Notably, the limitations in the uniqueness and computational efficiency of these methods become more significant in large-scale anonymization scenarios characterized by the generation of a huge number of pseudo-speakers, thus constraining their applications in the large-scale scenario.


This paper aims to improve the pseudo-speaker uniqueness and computational efficiency. A pseudo-speaker generator framework is proposed, which establishes a mapping from speaker identity index to speaker embedding vector with a feedforward network architecture, referred to as IDMap. During anonymization, the uniqueness of a newly generated pseudo-speaker is achieved by assigning an identity index randomly drawn without replacement. Specifically, the framework is specified with two models based on the realizations of the generator module, i.e., IDMap-MLP and IDMap-Diff. Between them, the IDMap-MLP utilizes a multi-layer perceptron (MLP) in the generator, and IDMap-Diff employs a diffusion network. Experimental evaluations were conducted on the small-scale LibriSpeech dataset\cite{panayotov2015librispeech} and the large-scale dataset comprising the MLS\cite{mls} and Common Voice\cite{commonvoice} datasets. The results demonstrate the effectiveness of the proposed IDMap framework in pseudo-speaker uniqueness, thereby improving the voice privacy protection capability, with enhanced computational efficiency. In the large-scale scenario, the proposed IDMap framework  further demonstrates improved stability in voice privacy protection as the number of generated pseudo-speakers increases.

The contributions of this paper include:

1. We introduce a feedforward framework, IDMap, for pseudo-speaker vector generation in voice anonymization. The framework is specified with two models: IDMap-MLP and IDMap-Diff. 

2. Both the IDMap-MLP and IDMap-Diff models were justified to increase the uniqueness of pseudo-speakers, thereby enhancing voice privacy protection capabilities. Moreover, the proposed models were validated for improved computational efficiency.

3. The efficacy of the proposed models was further validated in the large-scale scenario by demonstrating enhanced stability in voice privacy protection as the number of generated pseudo-speakers increased.

The remainder of this paper is organized as follows. Section \ref{sec.2} describes the voice anonymization framework used in our work. In Section \ref{sec:gan}, the GAN-based method is revisited, which generates the pseudo-speaker vector from a sampled vector. In Section \ref{sec.3}, the IDMap is illustrated, which generates the pseudo-speaker vector from a sampled speaker identity index. Experiments are presented in Section \ref{sec.exp} and conclusions are reached in Section \ref{sec.5}.

\begin{figure}[t]
  \centering 
  \includegraphics[scale=0.8]{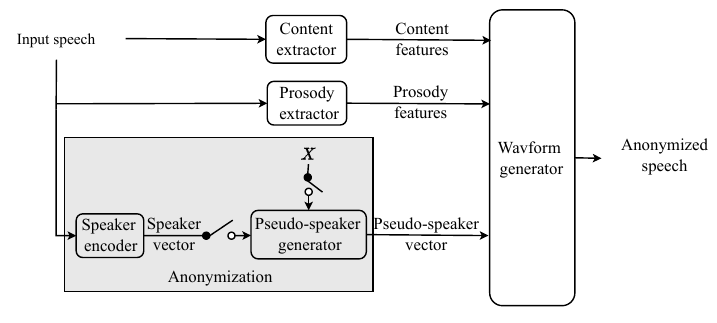}
  \caption{The voice anonymization framework. In the anonymization process, two types of variables can serve as inputs to different pseudo-speaker generator algorithms: the speaker vector extracted from the input utterance and a specific variable $X$.} 
  \label{fig:overview_framework} 
\end{figure}

\section{Overview of voice anonymization framework}
\label{sec.2}

Fig. \ref{fig:overview_framework} illustrates an overview of the voice anonymization framework, which is built upon a speech generation model wherein the linguistic content, speaker, and prosody attributes are disentangled and represented separately. As depicted in the figure, given an input speech, the content and prosody feature vectors are extracted using the respective content and prosody extractors. Meanwhile, the speaker embedding vector is extracted with a speaker encoder. An anonymization module is used to generate the speaker vector for the corresponding pseudo-speaker. Finally, a waveform generator is used to generate the anonymized speech given the content and prosody feature vectors extracted from the input utterance, along with the pseudo-speaker vector. Particularly, the waveform generator is trained using the content and prosody feature vectors extracted with the corresponding extractors, in conjunction with the speaker embedding vector derived from the speaker encoder.

As illustrated in Fig. \ref{fig:overview_framework}, the anonymization process is achieved using a pseudo-speaker generator, with distinct generator algorithms based on various input variables, primarily the speaker vector extracted from the input utterance and a specified variable $X$. Specifically, the generative methods, including the Gaussian mixture sampling \cite{turner2022generating} and GAN-based method \cite{GAN}, take both the original speaker vector and $X$, which is specified as a random seed, as input. Among the reference pool-based pseudo-speaker generator algorithms \cite{cohort-speaker}, the far and near proximity algorithms take both the original speaker vector and a random seed, represented by the input variable $X$, as input. The sparse, dense, and random proximity algorithms utilize a random seed $X$ as input. The transformation-based methods, such as the LOH \cite{ohnn} and SVD-based method\cite{svd}, take the original speaker vector as input. In our proposed pseudo-speaker generator based on the IDMap framework, the variable $X$ is specified as a sampled speaker identity index and applied as the input to the pseudo-speaker generator.

\section{Revisit of GAN-based method}
\label{sec:gan}

\begin{figure}[t]
    \hspace{0.02\textwidth}
    \begin{subfigure}[b]{0.21\textwidth} 
        \centering 
        \includegraphics[scale=0.9]{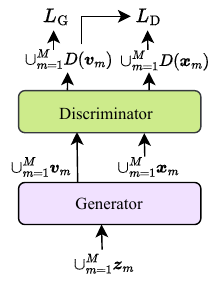}
        \caption{Training flow}
        \label{fig:GAN-based method training}
    \end{subfigure}
    \hspace{0.02\textwidth}
    \begin{subfigure}[b]{0.21\textwidth} 
        \centering
        \includegraphics[scale=0.9]{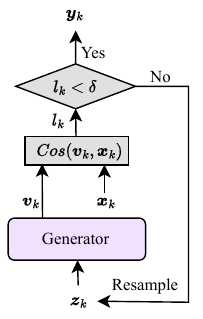}
        \caption{Inference flow}
        \label{fig:GAN-based method inference}
    \end{subfigure}
    \caption{The training and inference flows of the GAN-based pseudo-speaker generator.}
    \label{fig:GAN-based method}
\end{figure}

The GAN-based pseudo-speaker generator \cite{GAN} utilizes the GAN architecture using a Wasserstein GAN with quadratic transport cost (WGAN-QC)\cite{wgan-qc} as the loss function. The training process is illustrated in Fig. \ref{fig:GAN-based method}(\subref{fig:GAN-based method training}). Given the speaker vector extracted from the $i$-th speech utterance ${\boldsymbol{x}}_{i}$, a random vector \( \boldsymbol{z}_{i} \) is sampled from a stochastic distribution and then mapped to ${\boldsymbol{x}}_{i}$. Specifically, the generator generates \( \boldsymbol{v}_{i} \) from \( \boldsymbol{z}_{i} \), while the discriminator enforces distributional alignment between ${\boldsymbol{x}}_{i}$ and its prediction \( \boldsymbol{v}_{i} \).

The discriminator is optimized to approximate the quadratic
Wasserstein distance within each training mini-batch. Given a mini-batch of the speaker vectors extracted from $M$ utterances $\cup_{m=1}^M\boldsymbol{x}_{m}$ and the output vectors of the generator $\cup_{m=1}^M\boldsymbol{v}_{m}$, the loss function of the discriminator is formulated as:
\begin{equation}
\begin{split}
\mathcal{L}_{\text{D}} &= \frac{1}{2}\left(\frac{1}{m} \sum_{i=1}^m D(\boldsymbol{x}_{i})-\frac{1}{m} \sum_{i=1}^m H_{x,i}^{*}\right)^{2} \\
&+ \frac{1}{2}\left(\frac{1}{m} \sum_{i=1}^m\left(D(\boldsymbol{v}_{i})-H_{v,i}^{*}\right)^{2}\right) \\
&+ \frac{\gamma}{\sqrt{K} n} \sum_{i=1}^m \left(\left\| \nabla_{\boldsymbol{v}} D(\boldsymbol{v}_{i})\right\| -K\left\| \boldsymbol{x}_{i}-\boldsymbol{v}_{i}\right\| \right)^{2}
\end{split}
\label{eq:discriminator_loss}
\end{equation}
where the subscript $_{\rm D}$ is short for discriminator. $D(\bullet)$ is the discriminator neural network that takes real speaker vectors $\boldsymbol{x}$ or generated speaker vectors $\boldsymbol{v}$ as input. It outputs scores for each input vector. In (\ref{eq:discriminator_loss}), \( H_{x,i}^{*} \) and \( H_{v,i}^{*} \) are the optimal solutions for the real vector $\boldsymbol{x}_i$ and the generated vector $\boldsymbol{v}_i$, respectively, \( K \) is a positive real constant associated with the quadratic transport cost (set as \( 1/d \), \( d \) being the dimensionality of $\boldsymbol{x}$), \( \gamma \) is a weight variable, and $\nabla_{\boldsymbol{v}}$ is the derivative with respect to $\boldsymbol{v}$.

The generator loss is formulated as follows:
\begin{equation}
\mathcal{L}_{\text{G}} = -\,\frac{1}{m}\sum_{i=1}^m D\bigl(\boldsymbol{v}_{i}\bigr),
\label{eq:generator_loss}
\end{equation}
where the subscript \(_{\rm G}\) denotes the generator. Readers are referred to \cite{wgan-qc} for further details.

In its application in voice anonymization to generate the pseudo-speaker for the $k$-th original utterance, an input vector ${\boldsymbol{z}}_k$ is sampled first. Then the speaker vector ${\boldsymbol{v}}_k$ is generated from ${\boldsymbol{z}}_k$ by the generator. Thereafter, the cosine similarity $l_{k}$ between \({\boldsymbol{v}}_k\) and the speaker vector extracted from the original utterance, \({\boldsymbol{x}}_k\), is calculated and compared to a predefined threshold \(\delta\). As long as $l_{k}$ is higher than \(\delta\), ${\boldsymbol{z}}_k$ is resampled and a new ${\boldsymbol{v}}_k$ is generated. The process stops when the cosine similarity between \({\boldsymbol{v}}_k\) and \({\boldsymbol{x}}_k\) falls below \(\delta\). Finally, \({\boldsymbol{v}}_k\) is utilized as the pseudo-speaker vector for the utterance, denoted as ${\boldsymbol{y}}_k$. In our study, the uniform and Gaussian distributions are utilized for sampling the input vector \( \boldsymbol{z} \).

It is noteworthy that, when applied in the voice anonymization framework as illustrated in \textcolor{blue}{Fig. \ref{fig:overview_framework}}, the GAN-based pseudo-speaker generator takes the sampled vector \({\boldsymbol{z}}_k\) as the input variable \(X\) and transforms it into the speaker vector. Additionally, the original speaker vector ${\boldsymbol{x}}_k$ is utilized. As it is trained with the speaker vectors extracted from speech utterances, disregarding their speaker identities, it fails to ensure speaker distinctiveness among the generated speaker vectors, leading to a lack of uniqueness among the pseudo-speakers.

\section{IDMap}
\label{sec.3}

\begin{figure*}[t]
    \hspace{0.03\textwidth}
    \begin{subfigure}[b]{0.4\textwidth} 
        \centering 
        \includegraphics[scale=0.8]{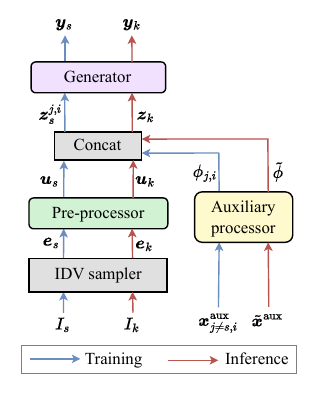}
        \caption{Overall architecture}
        \label{fig:anonymizer_overall}
    \end{subfigure}
    \hspace{0.015\textwidth}
    \begin{subfigure}[b]{0.21\textwidth} 
        \centering
        \includegraphics[scale=0.8]{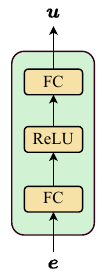}
        \caption{Pre-processor}
        \label{fig:uidv_processor}
    \end{subfigure}
    \hspace{0.02\textwidth}
    \begin{subfigure}[b]{0.21\textwidth} 
        \centering
        \includegraphics[scale=0.8]{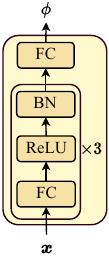}
        \caption{Auxiliary processor}
        \label{fig:identity_remover}
    \end{subfigure}
    \caption{The architecture of IDMap framework. FC is short for fully connected layer, and BN is short for batch normalization. In (a), the blue and red arrow lines are exclusively applicable during training and inference processes.}
    \label{fig:anonymizer}
\end{figure*}

Given a well-trained speech generation model, as illustrated in Fig. \ref{fig:overview_framework}, this paper proposes the IDMap framework for generating pseudo-speaker vectors. Based on the speaker embedding vector utilized in a specific speech generation model, IDMap establishes a mapping from speaker identity indices to speaker vectors. It is then utilized as the pseudo-speaker generator in the anonymization inference process. The overall architecture of the IDMap framework is presented in Fig. \ref{fig:anonymizer}.

\subsection{Overall framework}

Given the identity index of the $s$-th speaker, $I_{s} \left(I_{s}\in \mathbb{N}\right)$, and the corresponding speaker vector extracted from the speaker encoder applied in the speech generation model, ${\boldsymbol{x}}_s$, IDMap is trained to map $I_{s}$ to ${\boldsymbol{x}}_s$. As depicted in Fig. \ref{fig:anonymizer}(\subref{fig:anonymizer_overall}), given $I_s$, an identity vector (IDV) sampler is used to sample an identity vector from a stochastic distribution, yielding ${\boldsymbol{e}}_s$. Then, ${\boldsymbol{e}}_s$ goes through the pre-processor, obtaining an intermediate representation ${\boldsymbol{u}}_s$, which represents the speaker-specific identity information for speaker $s$. Simultaneously, an auxiliary speaker vector, ${\boldsymbol{x}}_{j \ne s, i}^{\text{aux}}$, extracted from the $i$-th utterance of speaker $j$, which is distinct from speaker $s$, is utilized. It is processed by the auxiliary processor, giving the vector $\phi_{j,i}$. Thereafter, ${{\boldsymbol{u}}}_s$ is concatenated with $\phi_{j,i}$ to be ${{\boldsymbol{z}}}_s^{j,i}$ and input into the generator, generating the predicted speaker vector ${\boldsymbol{y}}_{s}$. Finally, the framework is optimized under the supervision of ${\boldsymbol{x}}_s$.

In the training process, ${\boldsymbol{ x}}_{j \ne s, i}^{\text{aux}}$ is introduced to provide 1) data augmentation, 2) regularization, and 3) disturbance. Firstly, as the model is trained on speakers and the number of speakers available in the training dataset is always limited, the introduction of ${\boldsymbol{ x}}_{j \ne s, i}^{\text{aux}}$ to form a pair with \(I_{s}\) provides augmented training samples. Secondly, ${\boldsymbol{ x}}_{j \ne s, i}^{\text{aux}}$ provides auxiliary information about the intrinsic attributes of the speaker vector, thereby regularizing vector generation within the space of the speaker vectors. Lastly, the auxiliary speaker vector from speaker $j$, distinct from the training speaker $s$, introduces disturbance to the prediction of its speaker vector ${\boldsymbol{x}}_s$, thereby enhancing the capability of speaker vector generation. With this, the training sample of the IDMap framework is structured as a triplet $\left\{I_{s},{\boldsymbol{ x}}_s,{\boldsymbol{ x}}_{j \ne s, i}^{\rm{aux}}\right\}$. 
\begin{figure}[t]
  \centering 
  \includegraphics[scale=0.8]{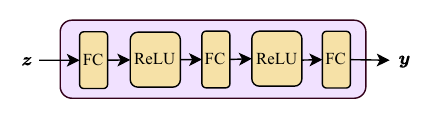}
  \caption{The generator in the IDMap-MLP model.} 
  \label{fig:MLP IDMap} 
\end{figure}

\subsubsection{IDV Sampler}  
Given a speaker identity index $I$, which is an integer, the IDV sampler samples a corresponding identity vector (IDV) \(\boldsymbol{e} \in \mathbb{R}^D\), where $D$ is the dimensionality of the identity vector. It leverages a permuted congruential generator (PCG) \cite{pcg} as a pseudorandom number generator (PRNG). Specifically, the PCG64 implementation from NumPy \cite{numpy} is utilized in our work. Given a random seed \(\omega\), a specified dimensionality \(D\), and a designated distribution $\mathcal{P}$, the algorithm generates a \(D\)-dimensional vector specific to \(\omega\), with each dimension following the distribution $\mathcal{P}$ independently. Based on that, the IDV sampler takes the identity index $I$ as the random seed \(\omega\) to generate a $D$-dimensional speaker vector ${\boldsymbol{e}}$. Two stochastic distributions are investigated in this paper for identity vector sampling: uniform and Gaussian distributions.

\subsubsection{Pre-processor}
The pre-processor transforms the identity vector ${\boldsymbol{e}}$ into the intermediate representation ${\boldsymbol{u}}$. As shown in Fig. \ref{fig:anonymizer}(\subref{fig:uidv_processor}), it consists of two fully connected layers with the ReLU activation function.

\subsubsection{Auxiliary processor}
As shown in Fig. \ref{fig:anonymizer}(\subref{fig:identity_remover}), in the auxiliary processor, a speaker vector ${\boldsymbol{x}}$ goes through three blocks of fully connected layers. In each block, the input vector passes through a fully connected layer, ReLU activation, and batch normalization successively. Thereafter, a fully connected layer is applied, outputting vector $\phi$.

\subsubsection{Generator}
The generator generates the predicted speaker vector given ${\boldsymbol y}_s$, ${\boldsymbol{ u}}_s$ and ${\phi}_{j, i}$. This paper investigates the implementation of the generator with a multi-layer perceptron (MLP) and diffusion probabilistic network, resulting in IDMap-MLP and IDMap-Diff models, respectively, detailed in the following.

\subsection{IDMap-MLP}
\label{sec:mlp}

The IDMap-MLP model utilizes an MLP as the generator, whose structure is shown in Fig. \ref{fig:MLP IDMap}. In the generator, the input vector ${\boldsymbol{z}}$ goes through two successive fully connected layers combined with ReLU activation, followed by a fully connected layer that produces the output vector ${\boldsymbol{y}}$.

The model is optimized using a loss function defined to maximize the speaker similarity and minimize the Euclidean distance between the ground-truth speaker vector ${\boldsymbol{x}}_s$ and its prediction ${\boldsymbol{y}}_s$, mathematically calculated as follows:

\begin{equation}
\mathcal{L}_{\text{MLP}} = \alpha \left( 1 - \frac{{\boldsymbol{x}}_s^{\mathsf T}{\boldsymbol{y}}_s}{\|{\boldsymbol{x}}_s\|_2\|{\boldsymbol{y}}_s\|_2} \right) + (1-\alpha) \|\boldsymbol{x}_s - \boldsymbol{y}_s\|_2,
\label{eq: regression loss}
\end{equation}
where $0\le\alpha\le1$ is the weight variable.

\begin{figure}[t]
  \centering
  \includegraphics[scale=0.7]{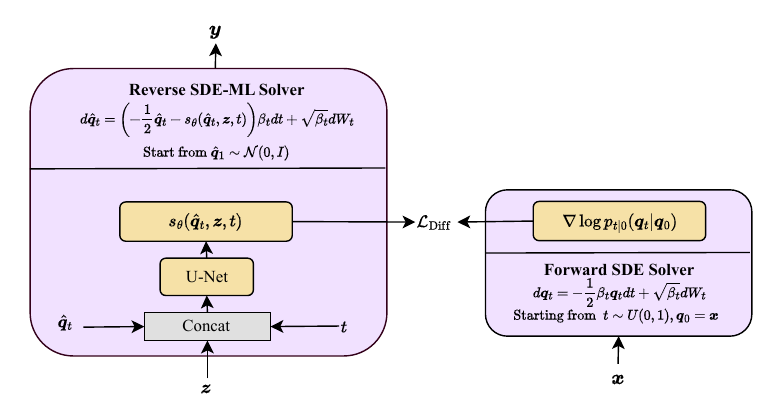}
  \caption{The forward and reverse process for the generator in the IDMap-Diff model.} 
  \label{fig:Diffusion IDMap} 
\end{figure} 

\subsection{IDMap-Diff}
\label{sec:diffusion}

The IDMap-Diff model employs a diffusion probabilistic network as the generator. The maximum likelihood stochastic differential equation solver (SDE-ML) proposed in \cite{DiffVC} is adopted, which derives discrete reverse steps to approximate the continuous diffusion process with minimal iterations for fast inference. The forward and reverse processes are illustrated in Fig. \ref{fig:Diffusion IDMap} and described in the following.

Given a ground-truth speaker vector $\boldsymbol{x}$ and the corresponding input vector ${\boldsymbol{z}}$ to the generator, the forward diffusion process gradually adds Gaussian noise according to a predefined schedule $\beta_t$ with the following SDE:   
\begin{equation}\label{eq:forward}
d\boldsymbol{q}_t = -\frac{1}{2} \beta_t \boldsymbol{q}_t dt + \sqrt{\beta_t} d\overrightarrow{W}_t,
\end{equation} 
where \( t \) is a continuous value within the interval \([0,1]\), representing the diffusion time step. When $t = 0$, $\boldsymbol{q}_0$ is initialized to $\boldsymbol{x}$, \( \beta_t \) is a noise schedule function, and \( d\overrightarrow{W}_t \) is a Wiener process. The forward SDE allows for an explicit solution: 
\begin{equation}\label{eq:noise}
\textit{p}(\boldsymbol{q}_t | \boldsymbol{q}_0) = \mathcal{N} \left( \gamma_{0,t} \boldsymbol{q}_0, \left(1 - \gamma_{0,t}^2 \right) I \right),
\end{equation}
where $\textit{p}(\boldsymbol{q}_t | \boldsymbol{q}_0)$ is the conditional probability density function of the final distribution of the forward process. $\gamma_{0,t}=e^{-\frac{1}{2}\int_0^t \beta_r dr}$, and $I$ is an $n\times n$ identity matrix. Meanwhile, the \( \beta_t\) follows a linear schedule \( \beta_t=\beta_0 + t(\beta_1 - \beta_0) \). When $t=1$, \(\gamma_{0,1}\) is close to zero, resulting in \( \textit{p}(\boldsymbol{q}_1) \) approaching \( \mathcal{N}(\boldsymbol{0}, I) \).

In the reverse process, the speaker vector $\boldsymbol{y}$ is predicted from noise $\hat{\boldsymbol{q}}_1$, which is sampled from the distribution defined in Eq. \eqref{eq:noise}. The reverse process is formulated as follows:
\begin{equation}
d\hat{\boldsymbol{q}}_t=\left(-\frac{1}{2}\hat{\boldsymbol{q}}_t - s_\theta(\hat{\boldsymbol{q}}_t, \boldsymbol{z}, t)\right)\beta_t dt+\sqrt{\beta_t}d\overleftarrow{W}_t,
\label{eq:reverse}
\end{equation}
where $s_\theta(\bullet)$ is the generator with the parameter set $\theta$, which estimates the gradient of the log-likelihood function of (\ref{eq:noise}). Additionally, $\boldsymbol{z}$ serves as an extra input to provide the specific speaker identity for the generated speaker vector $\boldsymbol{x}$, and $d\overleftarrow{W}_t$ is a Wiener process.

Given $\boldsymbol{z}$, the generator is trained to minimize the weighted mean square error (MSE) loss during the reverse diffusion, computed as follows:
\begin{equation}
\mathcal{L}_{\text{Diff}} = \int_0^1 \lambda_t \mathbb{E}_{\boldsymbol{q}_0, \boldsymbol{q}_t} \| s_\theta(\boldsymbol{q}_t,\boldsymbol{z}, t) - \nabla \log p_{t | 0}(\boldsymbol{q}_t | \boldsymbol{q}_0) \|_2^2 dt.
\label{eq: reverse process}
\end{equation}
In (\ref{eq: reverse process}), \( \lambda_t = 1 - e^{-\int_0^t \beta_r dr} \) is a weighting function, and \( \nabla \log p_{t | 0}(\boldsymbol{q}_t | \boldsymbol{q}_0) \) is the gradient of the log-likelihood function of (\ref{eq:noise}).

In our work, the U-Net architecture is applied in the diffusion probabilistic network. Readers are referred to \cite{DiffVC} for details.

\subsection{Inference and anonymization}
\label{sec:IDMap-infer}

Given the mapping from the speaker identity index to the speaker vector space established in the IDMap framework, the inference process is realized by sampling an identity index \( {I}_k \) from the input speaker identity indices and then mapping it to the speaker vector space. As shown in Fig. \ref{fig:anonymizer}(\subref{fig:anonymizer_overall}), in the inference process, given the speaker identity index $I_k$, an identity vector $\boldsymbol{e}_k$ is obtained with the IDV sampler first. Then, $\boldsymbol{e}_k$ goes through the pre-processor, giving ${\boldsymbol{u}}_k$. Meanwhile, an auxiliary speaker vector ${\tilde{{\boldsymbol{x}}}}^{\rm{aux}}$ is processed by the auxiliary processor, yielding ${\tilde{\phi}}$. The concatenation of ${\boldsymbol{u}}_k$ and ${\tilde{\phi}}$ is obtained as ${\boldsymbol{z}}_k$, which is then sent into the generator, generating the speaker vector ${\boldsymbol{y}}_k$. The auxiliary speaker vector ${\tilde{{\boldsymbol{x}}}}^{\text{aux}}$ is fixed and utilized for the generation of speaker vectors for any identity index. As justified in our experiment that will be presented in Section \ref{sec. x_aux justification}, ${\tilde{{\boldsymbol{x}}}}^{\rm{aux}}$ can be randomly selected from the training speaker vectors.

In its application in the voice anonymization framework as illustrated in Fig. \ref{fig:overview_framework}, IDMap takes a sampled speaker identity index \( I \) for the input $X$ and maps it to the corresponding speaker vector ${\boldsymbol{x}}$. Particularly, the previously generated speaker identity indices are stored in a set denoted as ${\mathcal I}$. Each time a new pseudo-speaker vector is generated, it is assigned a unique identity index that differs from those present in ${\mathcal I}$. This ensures that the newly generated speaker vector is distinct from all previously generated vectors, thereby achieving its uniqueness. Above all, denoting the model parameter set as $\theta$, the parameters of the anonyization process is summarized as $\left\{{\theta},{\mathcal I},\tilde{\boldsymbol{x}}^{\rm aux}\right\}$.

Notably, unlike the speaker vectors extracted from utterances used to train the GAN-based model, IDMap is trained with speaker vectors derived from speakers, ensuring voice distinctiveness among the generated speaker vectors.


\begin{figure}[t]
  \centering
  \includegraphics[scale=0.7]{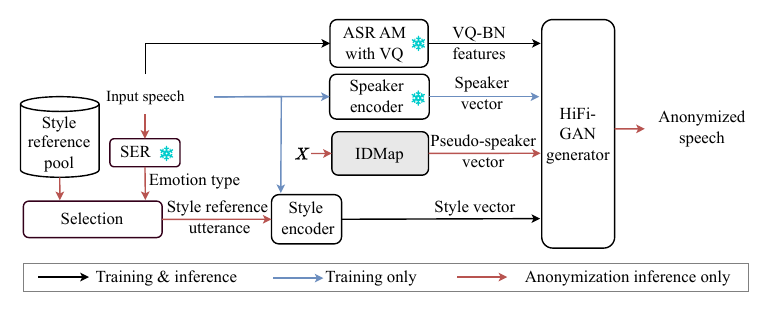}
  \caption{The inference flows of the voice anonymization framework applied in our work. The ASR AM with VQ extractor, speaker encoder, and the SER model are pre-trained and frozen. The IDMap takes a sampled input identity index as $X$. The black lines are valid in training and anonymization inference, while the blue lines are applicable only in training, and the red lines are applicable only in anonymization inference.}
  \label{fig:specific_framework} 
\end{figure}

\section{Application in anonymization}

Fig. \ref{fig:specific_framework} presents the specific speech generation model utilized in this study, which is derived from that proposed in \cite{vq-gst}. Given an input speech utterance, the linguistic content is represented by the bottleneck (BN) feature vectors extracted using an ASR AM with a vector quantization (VQ) extractor \cite{sun2016phonetic}, yielding the VQ-BN feature vectors. The speaker attribute is extracted using a speaker encoder and represented as a speaker embedding vector \cite{snyder17_interspeech}. The style embedding vector is extracted via a style encoder, which utilizes the non-content learner structure in \cite{vq-gst}. Specifically, it is based on the global style token (GST) mechanism \cite{wang2018style}. The VQ-BN feature vectors, speaker vector, and style vector are fed into the HiFi-GAN generator \cite{kong2020hifi} to generate the output speech. In the framework, the ASR AM with VQ extractor and the speaker encoder are pre-trained, while the style encoder and HiFi-GAN generator are jointly trained.

During anonymization, given an input speech utterance, the VQ-BN feature vectors are extracted. A well-trained speech emotion recognition (SER) model is utilized to identify the emotion type of the original speech, followed by the selection of a style reference utterance exhibiting the same emotional state from the style reference pool. Thereafter, the style embedding vector is extracted from the style reference utterance. The corresponding pseudo-speaker vector is obtained by the IDMap models, which takes a sampled identity index as input, denoted as $X$. Finally, the VQ-BN feature vectors, the pseudo-speaker vector, and the style vector are input into the HiFi-GAN generator \cite{kong2020hifi} to generate the anonymized speech.

\section{Experiments}
\label{sec.exp}

\subsection{Evaluation metrics}
Our experiments were carried out following the configurations provided by VPC2024\cite{vpc2024}\footnote{\label{vpc2024_eval}\url{https://github.com/Voice-Privacy-Challenge/Voice-Privacy-Challenge-2024}}. Automatic speaker verification (ASV) tests were conducted to evaluate the voice protection capability, measured by equal error rates (EERs). The linguistic content preservation capability was examined in automatic speech recognition (ASR) tests, measured by word error rates (WERs). The ability to preserve the emotional state of the original utterance was assessed with speech emotion recognition (SER), measured by unweighted average recall (UAR).

\subsection{Datasets}
In our experiments, the LibriTTS train-clean-100, train-clean-360, and train-other-500 subsets \cite{libritts} were used for training the speech generation model depicted in Fig. \ref{fig:specific_framework}. The LibriTTS train-clean-100 subset was employed to train the proposed IDMap-MLP and IDMap-Diff pseudo-speaker generators, consisting of \text{33,236} speech utterances from \text{247} speakers.
During anonymization, the ESD dataset\cite{esd} was used for the style reference pool, which comprises 350 utterances originating from 10 speakers across five emotions: neutral, happy, angry, sad, and surprised. Following the VPC 2024 setting, ``surprise'' and ``happy'' were merged into one emotion class, resulting in four emotions. Evaluations were performed in both small- and large-scale scenarios. The development and test subsets of LibriSpeech \cite{panayotov2015librispeech} were utilized for the small-scale evaluation. The MLS \cite{mls} and the Common Voice datasets \cite{commonvoice} were used for the large-scale evaluations. Besides, the IEMOCAP \cite{IEMOCAP} dataset was used for emotion preservation evaluation, with the development and test subsets constructed following the VPC2024 configurations \cite{vpc2024}. The LibriSpeech train-clean-360 dataset was used for training the ASV evaluation models.

\subsection{Speech generation model}
In the speech generation model as depicted in Fig. \ref{fig:specific_framework}, the PPG extractor, speaker encoder, and SER model were pre-trained. The PPG extractor adopted the VQ-BN extractor as proposed in \cite{vq-bn}, trained with the open-source code \footnote{\url{https://github.com/deep-privacy/SA-toolkit/tree/master/egs/asr/librispeech}}. The publicly available wav2vec2.0\footnote{\url{https://dl.fbaipublicfiles.com/fairseq/wav2vec/w2v_large_lv_fsh_swbd_cv_ftls960_updated.pt}} was used in the VQ-BN extractor which was pre-trained on Libri-Light \cite{librilight}, CommonVoice \cite{commonvoice}, Switchboard \cite{switchboard}, and Fisher datasets \cite{fisher}, and then fine-tuned with the Librispeech dataset. In its application in the speech generation model, the VQ-BN extractor was trained with the Librispeech train-clean-100 dataset with the wav2vec2.0 module frozen. The speaker encoder utilized the ECAPA-TDNN encoder architecture\cite{ECAPATDNN}. The pooling algorithm adopted the xi-vector \cite{xivector} strategy, yielding a 512-dimensional speaker mean vector combined with uncertainty. The mean vector was used as the speaker vector. The speaker extractor was trained on the VoxCeleb 1 \& 2 datasets \cite{vox1,vox2}, using the open-source ASV-Subtools toolkit\footnote{\url{https://github.com/Snowdar/asv-subtools/blob/master/pytorch/launcher/r
unEcapaXvector online.py}}. Given the pre-trained VQ-BN extractor and speaker encoder, the style encoder and HiFi-GAN generator within the speech generation model were trained jointly on Librispeech train-clean-100, train-clean360, and train-other-500 datasets. The style encoder was composed of 10 SE-ResNet \cite{se-resnet} layers followed by a gated recurrent unit (GRU) layer, and employed 8 style tokens. The HiFi-GAN generator adopted the architecture proposed in \cite{kong2020hifi}. In anonymization inference, an SER model with the architecture presented in \cite{mmer} was applied. It was trained on ESD  with the open-source code \footnote{\url{https://github.com/Sreyan88/MMER?tab=readme-ov-file}}.

\subsection{Compared methods}
Four baseline methods were compared with the proposed IDMap-MLP and IDMap-Diff pseudo-speaker generators in our experiments. The baseline methods include random selection, average, pseudo-speaker distribution, and GAN-based approaches. The details of the compared methods are as follows.

\noindent \romannumeral 1. \emph{Random selection (RS)}: A speaker was randomly selected from the reference pool as the pseudo-speaker.

\noindent \romannumeral 2. \emph{Average}\cite{cohort-speaker}: Given a reference pool, \text{100} cohort speakers with the furthest distances from the original speaker were selected from the reference pool. The pseudo-speaker vector was obtained as the average of the speaker vectors of the cohort speakers.

\noindent \romannumeral 3. \emph{Pseudo-speaker distribution (PSD)}\cite{psd}: Given a reference pool, the cohort speakers were selected according to the dense proximity\cite{cohort-speaker}. Speaker distributions, parameterized by mean and uncertainty, were estimated within the speech frames of the cohort utterances using the speaker encoder. The pseudo-speaker distributions were estimated from these frame-level speaker distributions.

\noindent \romannumeral 4. \emph{GAN}-based\cite{GAN}:  A GAN-based pseudo-speaker generator was trained to generate pseudo-speakers from randomly sampled input vectors. The generator and discriminator shared the same structure, consisting of 3 three residual blocks as used in \cite{wgan-qc}. The generator takes a 512-dimensional input vector and outputs a 512-dimensional speaker vector. The LibriTTS train-clean-100 dataset was used for training with a batch size of 64. In inference, the cosine similarity threshold $\delta$ was set to 0.3.
 
\noindent \romannumeral 5. \emph{IDMap-MLP}: In the experiments of IDMap-MLP, the IDV sampler generated 512-dimensional identity vectors. The layer sizes in the pre-processor were \text{512-512-512}. In the auxiliary processor, the layer sizes of a block were \text{512-512}, and the output layer size was \text{512}. By concatenating the outputs of the pre-processor and the auxiliary processor, the input to the generator was of 1024 dimensions. The layer sizes of the generator were 1024-512-512. In the loss function (\ref{eq: regression loss}), $\alpha$ was set to 0.5.

\noindent \romannumeral 6. \emph{IDMap-Diff}: In the IDMap-Diff model, the same IDV sampler, pre-processor, and auxiliary processor utilized in the IDMap-MLP were applied. In the generator, the U-Net architecture employed in the DiffVC model \cite{DiffVC} was utilized, implemented using the source code available at \footnote{\url{https://github.com/agoyr/DiffVC}}. Specifically, three feature map resolutions were used in the U-Net with an
additional channel added for the input $\boldsymbol{z}$. A 5-step SDE was used with linear noise schedule \( \beta_t = \beta_0 + t(\beta_1 - \beta_0) \), where $\beta_0 = 0.05$ and $\beta_1 = 20.0$. The input channel was set to 1024 while the output channel was set to 512.

Both the IDMap-MLP and IDMap-Diff models were trained with LibriTTS train-clean-100 \cite{libritts} in the following configurations. In each training mini-batch, 16 speaker identity indices were included. For each training speaker, the speaker vector ${\boldsymbol{x}}$ was obtained by averaging those extracted from her/his utterances. Meanwhile, 16 auxiliary utterances were randomly selected from speakers distinct from the training speaker, from which the auxiliary speaker vectors were extracted. This process yielded 256 training triplets of $\{I_s, \boldsymbol{x}_s, \boldsymbol{x}_{j \ne s, i}^{\text{aux}}\}$ per mini-batch. In anonymization inference, $\tilde{{\boldsymbol{x}}}^{\rm aux}$ was randomly selected from the training set.

The random selection, average, and PSD methods are reference pool-based methods, with the LibriTTS train-clean-100 dataset used as the reference pool for cohort speaker selection. Both the uniform distribution $\mathcal{U}\left(-1, 1\right)$ and the standard normal distribution $\mathcal{N}\left(0, 1\right)$ were applied for input vector sampling in the GAN-based method and for IDV sampling in the IDMap specifications. Mean variance normalization (MVN) was applied to normalize the sampled vectors in these methods.

\begin{table*}[t]
    \centering
    \caption{Performances of compared anonymization methods, including the EERs (\%), WERs (\%), and UARs (\%) obtained in the ASV, ASR, and SER evaluations. EERs are presented for the development (dev) and test subsets of LibriSpeech (libri), and for male (m) and female (f) genders, respectively. The average EERs obtained across the evaluation subsets are presented in the row of \emph{avg} for each method. The comparison between the baseline methods of random selection (RS), Average, PSD, and GAN-based methods, and the proposed IDMap-MLP, IDMap-Diff models is presented. Results obtained using both uniform ($\mathcal U$) and standard normal ($\mathcal N$) distributions for input vector sampling in the GAN-based methods, along with identity vector sampling in the proposed IDMap-MLP and IDMap-Diff models, are included.}
    \label{tab:E-WER}
        \begin{tabular}{c|c|c|c|c|c|cc|cc|cc}
            \Xhline{1px}
            \multirow{2}{*}{} & 
            \multirow{2}{*}{Dataset} & 
            \multirow{2}{*}{Gender} & 
            \multirow{2}{*}{RS} & 
            \multirow{2}{*}{Average} & 
            \multirow{2}{*}{PSD} & 
            \multicolumn{2}{c|}{GAN} & 
            \multicolumn{2}{c|}{IDMap-MLP} & 
            \multicolumn{2}{c}{IDMap-Diff} \\
            \cline{7-12}
            & & & & & & $\mathcal{N}$ & $\mathcal{U}$ & $\mathcal{N}$ & $\mathcal{U}$ & $\mathcal{N}$ & $\mathcal{U}$ \\
            \Xhline{1px}
            \multirow{6}{*}{EER} 
            & \multirow{2}{*}{libri-dev} & f 
            & 41.47 & 41.79 & 45.49 & 42.47 & 43.59 & 46.16 & 46.32 & \textbf{48.47} & 48.21 \\
            \cline{3-12}
            & & m 
            & 43.22 & 39.60 & 44.31 & 44.22 & 40.77 & 46.60 & 44.24 & \textbf{47.86} & 47.36 \\
            \cline{2-12}
            & \multirow{2}{*}{libri-test} & f 
            & 41.54 & 39.43 & 43.43 & 43.54 & 44.62 & 43.61 & 43.81 & \textbf{48.18} & 47.66 \\
            \cline{3-12}
            & & m 
            & 40.58 & 39.87 & 42.29 & 40.60 & 41.99 & 45.83 & 46.32 & \textbf{48.46} & 47.34 \\
            \cline{2-12}
            & \multicolumn{2}{c|}{avg} 
            & 41.70 & 40.17 & 43.88 & 42.70 & 42.74 & 45.54 & 45.17 & \textbf{48.24} & 47.64 \\
            \hline\hline
            \multirow{2}{*}{WER} 
            & libri-dev & - 
            & 3.31 & 3.37 & 3.39 & 3.38 & 3.45 & 3.37 & 3.38 & 3.38 & 3.41 \\
            \cline{2-12}
            & libri-test & - 
            & 3.21 & 3.23 & 3.23 & 3.28 & 3.23 & 3.23 & 3.25 & 3.22 & 3.28 \\
            \hline\hline
            \multirow{2}{*}{SER} 
            & IEMOCAP-dev & - 
            & 53.47 & 52.25 & 53.42 & 53.48 & 52.74 & 53.67 & 52.79 & 53.78 & 54.23 \\
            \cline{2-12}
            & IEMOCAP-test & - 
            & 52.21 & 52.11 & 53.88 & 53.43 & 53.23 & 52.63 & 53.88 & 52.85 & 53.01 \\
            \Xhline{1px}
        \end{tabular}
\end{table*}

\begin{table*}[t]
    \centering
    \caption{$G_{\rm vd}$ (dB) and DeID (\%) results on pooled development and test subsets of LibriSpeech. The random selection (RS), average, PSD, GAN-based, and the proposed IDMap-MLP and IDMap-Diff methods are included. In the GAN-based, IDMap-MLP, and IDMap-Diff methods, the results obtained by employing both the uniform (\(\mathcal{U}\)) and standard normal (\(\mathcal{N}\)) distributions for vector sampling are presented.}
    \begin{tabular}{c|c|c|c|cc|cc|cc}
        \Xhline{1pt}
        \multirow{2}{*}{} & 
        \multirow{2}{*}{RS} & 
        \multirow{2}{*}{Average} & 
        \multirow{2}{*}{PSD} & 
        \multicolumn{2}{c|}{GAN} & 
        \multicolumn{2}{c|}{IDMap-MLP} & 
        \multicolumn{2}{c}{IDMap-Diff} \\
        \cline{5-10}
        & & & & $\mathcal{N}$ & $\mathcal{U}$ & $\mathcal{N}$ & $\mathcal{U}$ & $\mathcal{N}$ & $\mathcal{U}$ \\
        \hline
        $G_{\rm vd}$ & 
        -1.684 & 
        -3.056 & 
        0.387 & 
        -0.137 & 
        -0.144 & 
        0.412 & 
        0.406 & 
        0.521 & 
        0.513 \\
        \hline
        DeID & 
        98.43 & 
        98.23 & 
        99.24 & 
        98.93 & 
        98.75 & 
        99.49 & 
        99.31 & 
        99.76 & 
        99.96 \\
        \Xhline{1pt}
    \end{tabular}
    \label{tb: GVD DeID}
\end{table*}

\subsection{Voice anonymization evaluations}
\label{sec:eva}
Voice anonymization evaluations were conducted following the VPC2024 configurations and implemented using the open-source recipe \footref{vpc2024_eval}. In our experiments, the methods were first examined on the small-scale dataset, composed of the development and test subsets of LibriSpeech. ASV, ASR, and SER evaluations were performed, as detailed below.

\subsubsection{ASV evaluations}
\label{sec.vpc2024_asv}
In the ASV evaluations for each method, evaluation models were trained using the anonymized speech utterances from the LibriSpeech train-clean-360 dataset. The utterances used for model training and evaluation were anonymized at the utterance level, i.e., generating a pseudo-speaker for each utterance. Following the VPC2024\footref{vpc2024_eval} recipe, an ECAPA-TDNN model was trained to extract speaker vectors, combined with a probabilistic linear discriminant analysis (PLDA) backend for scoring. The ASV evaluations were conducted in a gender-dependent manner. The EERs obtained on the female and male utterances of the development and test datasets of LibriSpeech are presented in Table \ref{tab:E-WER}, along with the average across the four subsets for each method. For the GAN-based and proposed IDMap-MLP and IDMap-Diff methods, the results are presented in terms of both uniform and standard normal distributions for vector sampling.

From the table, it can be observed that the proposed IDMap-Diff model achieved the highest EERs across all compared methods, and IDMap-MLP achieved the second highest, demonstrating the superiority of the proposed IDMap framework in voice privacy protection. Compared to the baseline methods, including random selection, averaging, PSD, and GAN-based approaches, the superiority of the proposed IDMap is attributed to its mechanism for unique pseudo-speaker generation, achieved through a mapping from speaker identity index to speaker vector. Especially, compared to the GAN-based method, which generates speaker vectors from sampled vectors similarly to IDMap, the superiority of IDMap should be due to its training with speaker-discriminative speaker vectors, leading to enhanced voice distinctiveness among the generated speaker vectors. Furthermore, IDMap-Diff achieved higher EERs than IDMap-MLP, indicating the advantage of the diffusion network as a generator. This should be because the diffusion network has greater capability in vector generation than the simple MLP, thereby better representing the discrimination among the speaker identity indices.

\subsubsection{ASR evaluations}
Following the VPC2024 recipe\footref{vpc2024_eval}, the ASR evaluation model adopted the wav2vec 2.0 architecture, which was fine-tuned on the combination of train-clean-100, train-clean-360, and train-other-500 subsets of LibriSpeech. The evaluations were carried out on the development and test subsets of the LibriSpeech dataset. The WERs obtained by the compared methods are presented in Table \ref{tab:E-WER}. From the table, it can be observed that the compared methods achieved similar WERs. This suggests that the proposed IDMap framework, in both the IDMap-MLP and IDMap-Diff models, did not degrade the linguistic content preservation capability within the voice anonymization framework.

\subsubsection{SER evaluations}
Following the VPC2024 recipe\footref{vpc2024_eval}, the SER model was trained on the training subset of the IEMOCAP dataset. The UARs obtained on the development and test sets of IEMOCAP by the compared methods are presented in Table \ref{tab:E-WER}. From the results, it is shown that the proposed IDMap-MLP and IDMap-Diff models achieved comparable UARs with the baseline methods. This indicates that the proposed IDMap framework is capable of preserving the emotion of the original utterance within the voice anonymization framework.


\subsection{Gain of voice distinctness ($G_{\rm vd}$) evaluations} 
$G_{\rm vd}$ values \cite{noe20_interspeech} were computed to measure the uniqueness of the pseudo-speakers. In these tests, the evaluation utterances were anonymized at the speaker level, with a unique pseudo-speaker generated for each original speaker and applied to all utterances from that speaker. The evaluations were performed on the pooled development and test subsets of the LibriSpeech dataset, with results presented in Table \ref{tb: GVD DeID}. From the results, it can be observed that IDMap-MLP and IDMap-Diff models obtained higher $G_{\text{vd}}$ values than the baseline methods, further justifying the superiority of the proposed IDMap framework in generating unique pseudo-speakers. Moreover, compared to IDMap-MLP, IDMap-Diff achieved even higher \(G_{\text{vd}}\) values, demonstrating that the diffusion-based pseudo-speaker generator provided a stronger generation capability. The observations from the \(G_{\text{vd}}\) comparison further substantiate the advantages of IDMap over baseline methods in voice privacy protection, evaluated in the ASV tests, by achieving improved pseudo-speaker uniqueness.

\subsection{De-identiﬁcation (DeID) evaluations} 
DeID evaluations \cite{noe20_interspeech} were conducted to measure the effectiveness of de-identification of the anonymized speech utterances. Like the $G_{\rm vd}$ evaluation, the evaluation utterances were anonymized at the speaker level and the evaluations were performed on the pooled development and test subsets of the LibriSpeech dataset. The results obtained on the compared methods are presented in Table \ref{tb: GVD DeID}. From the results, it can be observed that all methods achieved high DeID values, close to 100\%. Moreover, the proposed IDMap-MLP and IDMap-Diff models outperformed the baseline methods, demonstrating superior de-identification efficacy.

\begin{table}[t]
    \centering
    \caption{RTFs obtained by the compared methods including average, PSD, GAN, IDMap-MLP, and IDMap-Diff.}
    \label{tab:RTF}
    \begin{tabular}{c|c|c|c|c}
    \Xhline{1pt}
    Average & PSD & GAN  & IDMap-MLP & IDMap-Diff  \\ \hline
    $2.113 \times 10^{-4}$ & 3.384 & 0.0419 & $6.417 \times 10^{-4}$ & $2.43 \times 10^{-3}$ \\ \Xhline{1pt}
    \end{tabular}
\end{table}

\subsection{Computational efficiency}

The computational efficiencies of the compared methods were measured with real-time factor (RTF). The RTF was computed as the ratio between the time cost of generating speaker vectors and the input audio duration. In this evaluation, 1000 utterances were randomly selected from the development and test subsets of the LibriSpeech dataset. For a fair comparison, the standard normal distribution $\mathcal{N}(0, 1)$ was applied for vector sampling in the GAN-based, IDMap-MLP, and IDMap-Diff models. Since the random selection method does not involve generating speaker vectors, it was excluded from the RTF evaluations. The results are presented in Table \ref{tab:RTF}. From the results, it can be observed that the average method achieved the highest computational efficiency as it requires only a simple averaging operation on the cohort speaker vectors. Besides, the proposed IDMap-MLP and IDMap-Diff models obtained superior efficiency than the GAN-based and PSD methods. Moreover, between the IDMap-MLP and IDMap-Diff models, the latter required a higher time cost than the former due to its iterative implementation in inference. However, its time cost remained significantly lower than those of the examined model-based methods, wherein models were utilized to generate pseudo-speaker vectors, including the PSD and GAN-based approaches. This indicates enhanced computational efficiency of the proposed IDMap framework.

\subsection{Supplementary evaluations}

\subsubsection{Comparison between $\mathcal{L}_{\rm MLP}$ and WGAN-QC losses in IDMap-MLP}
\begin{figure}[t]
  \centering
  \includegraphics[width=0.99\columnwidth]{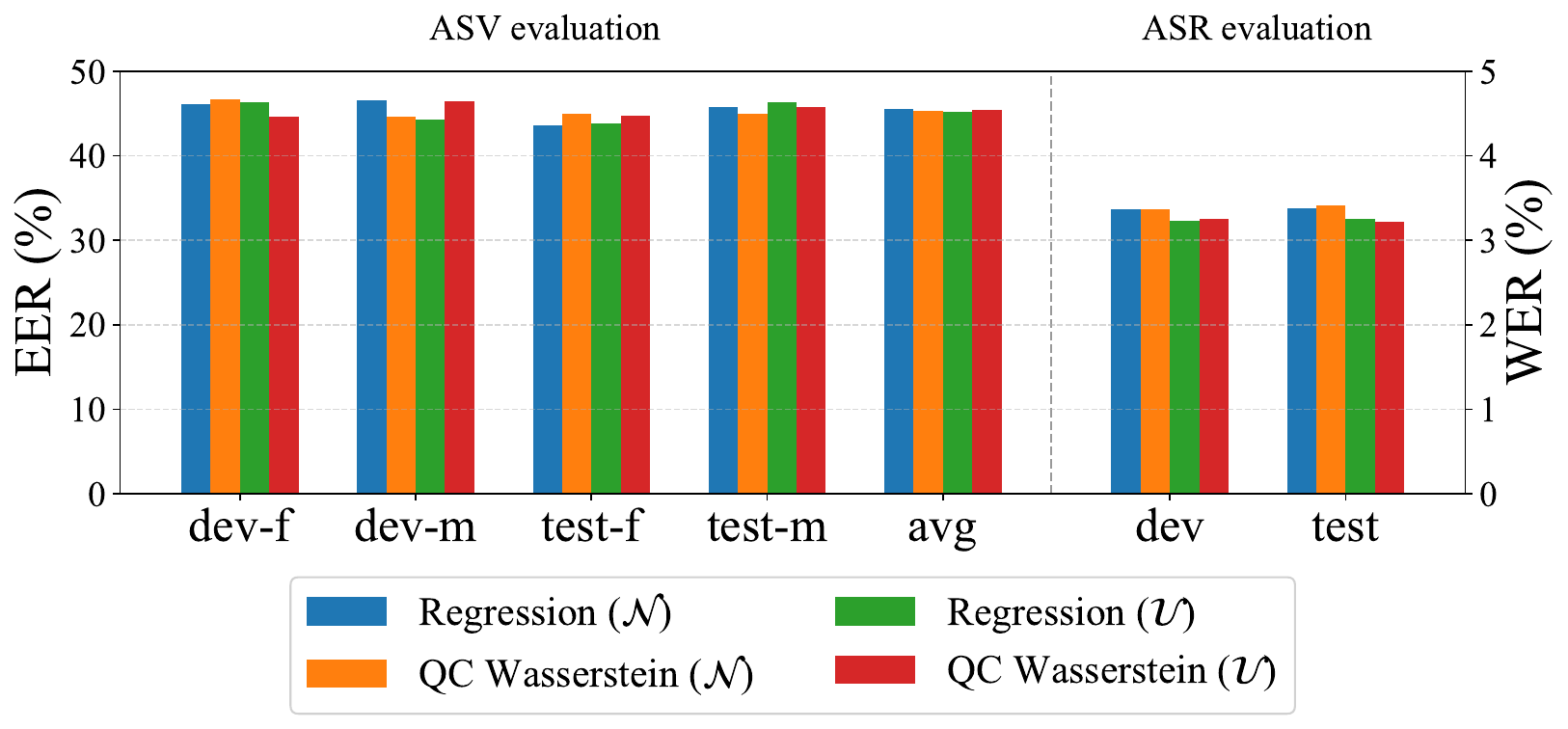}
  \caption{Performance comparison between ${\mathcal L}_{\rm MLP}$ and WGAN-QC loss functions in IDMap-MLP. Both EERs (\%) and WERs (\%) obtained in the ASV and ASR evaluations on the development (dev) and test subsets of LibriSpeech are presented. Results obtained by using both uniform ($\mathcal{U}$) and standard normal ($\mathcal{N}$) distributions for identity vector sampling are included. Separated by the dotted line, the left side presents the EERs, the right side shows the WERs. The EERs are given in a gender-independent manner for male (m) and female (f), respectively. The average EERs obtained across the subsets for the compared configurations are presented in \emph{avg}.} 
  \label{fig:ablation-loss}
\end{figure}

Since both the WGAN-QC loss used in the GAN-based method and ${\mathcal L}_{\rm MLP}$ in the proposed IDMap-MLP model can be applied in speaker vector generation, experiments were conducted to compare these two loss functions in the IDMap-MLP model. In this comparison, the IDMap-MLP model was trained with the WGAN-QC loss and \(\mathcal{L}_{\text{MLP}}\), respectively. In the application of the WGAN-QC loss in the IDMap-MLP model, the same discriminator architecture used in the GAN-based model was applied. Experiments were conducted using both the uniform and standard normal distributions for the IDV sampling. The experimental results obtained by the two loss functions in ASV and ASR evaluations on the development and test subsets of LibriSpeech are presented in Fig. \ref{fig:ablation-loss}. In the EER comparison, the average EER obtained across the four evaluation subsets for each method is presented. The comparison indicates that, under both uniform and standard normal distributions for IDV sampling, \(\mathcal{L}_{\text{MLP}}\) achieved performance comparable to the WGAN-QC loss in both ASV and ASR evaluations. Moreover, compared with the WGAN-QC loss function which involves an additional discriminator, \(\mathcal{L}_{\text{MLP}}\) achieves similar performance with lower implementation complexity.

\subsubsection{Ablation study in ${\mathcal L}_{\rm MLP}$}

\begin{figure}[t]
  \centering
  \vspace{0.265cm}  \includegraphics[width=0.99\columnwidth]{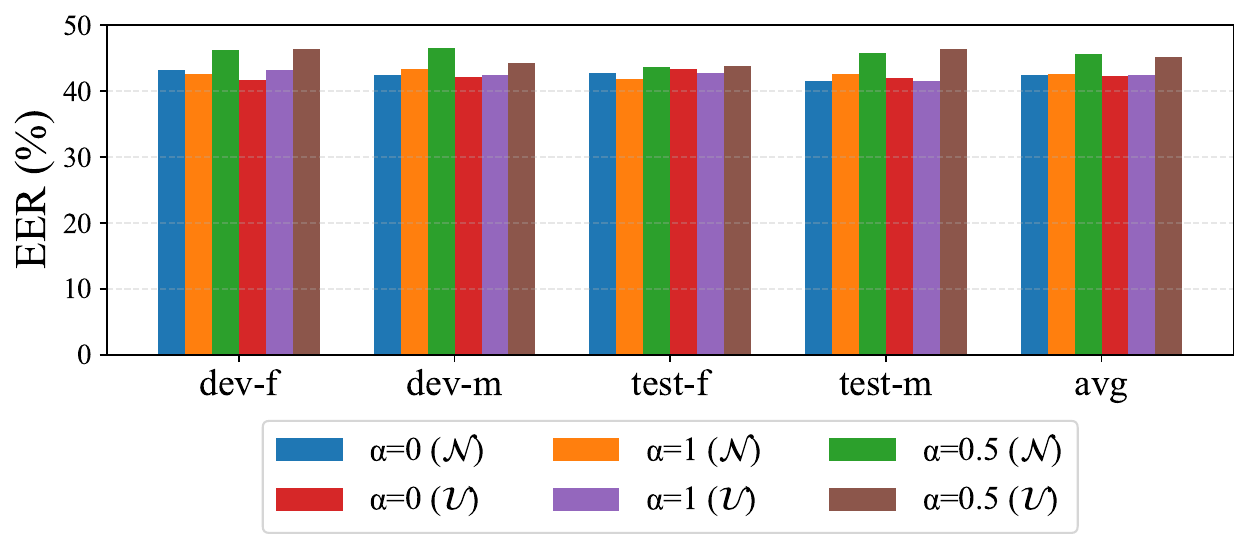}
  \caption{EERs(\%) obtained by different $\alpha$ configurations in $L_{\rm MLP}$ by setting $\alpha \in \left\{0,0.5,1\right\}$. Results obtained on the development and test subsets of LibriSpeech are presented separately in males (m) and females (f). Results obtained by applying both uniform ($\mathcal{U}$) and standard normal ($\mathcal{N}$) distributions for identity vector sampling are included. EER values are averaged across the datasets for each configuration and presented in \emph{avg}.} 
  \label{fig:ablation-loss-a}
\end{figure}

To validate the effectiveness of the two terms in ${\mathcal L}_{\rm MLP}$, ablation studies were conducted by setting $\alpha$ to 0 and 1, respectively. Specifically, by setting \(\alpha=0\), the cosine similarity was excluded from the loss function, whereas setting \(\alpha=1\) resulted in the exclusion of the Euclidean distance. The results obtained by the configurations utilizing uniform and standard normal distributions for IDV sampling are presented in Fig. \ref{fig:ablation-loss-a}. From the figure, it can be observed that setting \(\alpha = 0\) or \(\alpha = 1\) decreased the EERs in the ASV evaluations compared to \(\alpha = 0.5\), thereby validating the effectiveness of both the cosine similarity and Euclidean distance terms in ${\mathcal L}_{\rm MLP}$.

\subsubsection{Regularization function of ${\boldsymbol{x}}^{\rm aux}$}
\label{sec. x_aux justification}
Next, experiments were conducted on both the IDMap-MLP and IDMap-Diff models to examine the function of the auxiliary speaker vector ${\boldsymbol{x}}^{\rm aux}$ in regularization. The models trained with the standard normal distribution for identity vector sampling were examined. Given a speaker vector ${\boldsymbol{x}}$ extracted from an utterance, after going through the first two blocks and the output layer in the auxiliary processor, the vectors were obtained and represented as $\phi^{\rm 1}$, $\phi^{\rm 2}$, and $\phi$, respectively. For each speaker, the original speaker vector \( \boldsymbol{x} \), \( \phi^{\text{1}} \), \( \phi^{\text{2}} \), and \( \phi \) were obtained by averaging the corresponding vectors derived from all of their speech utterances. For each vector type, cosine similarity was calculated between each speaker and all other speakers in the dataset. Then the average of the similarities was calculated for each vector type across all the speaker pairs. The results obtained on the LibriSpeech train-other-500 dataset are given in Table \ref{tab:intra_class_similarity}. The higher the value, the more speaker-specific information was contained in the vector.

\begin{figure*}[t]
  \centering 
  \includegraphics[scale=0.5]{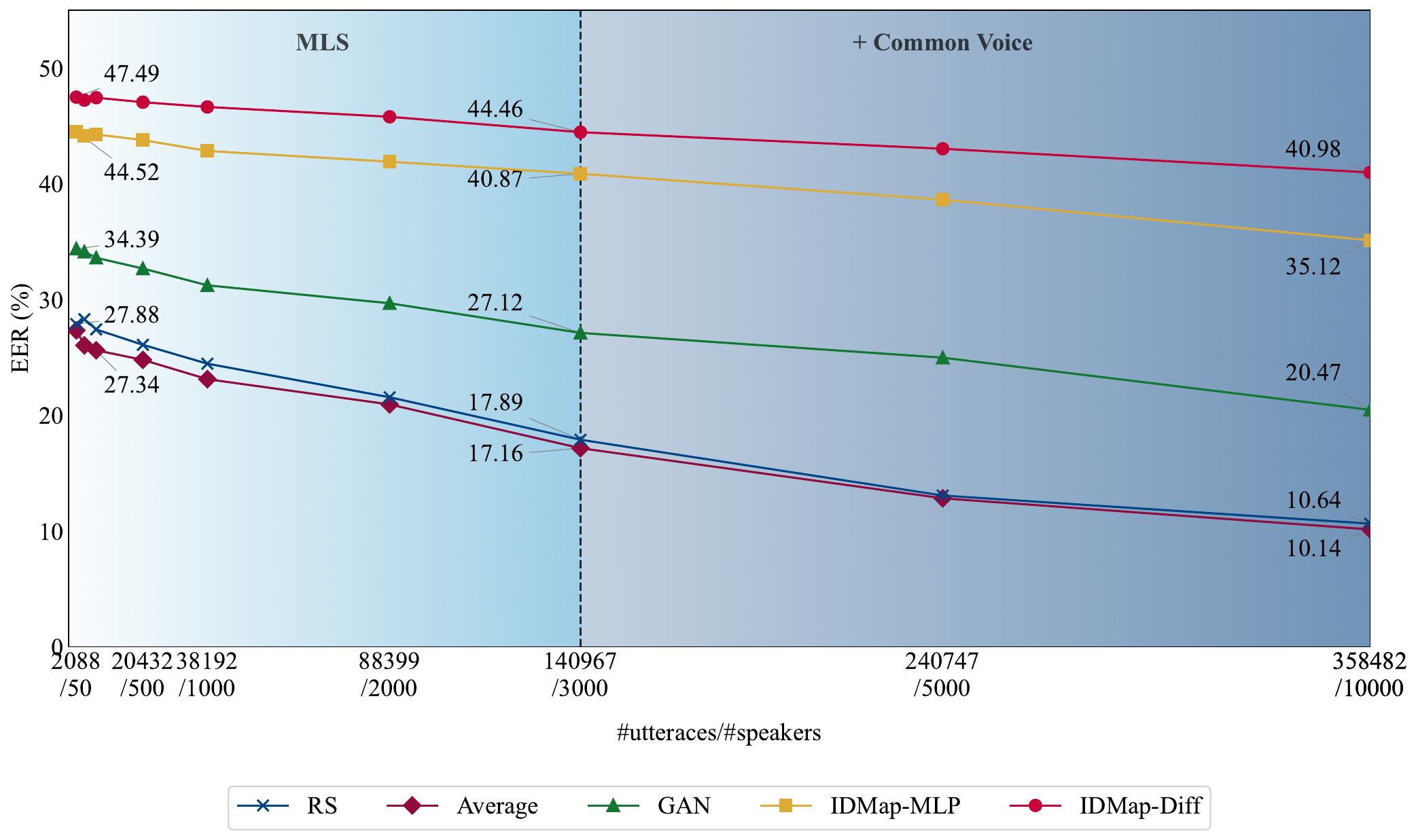}
  \caption{The EERs (\%) achieved by the compared methods of random selection (RS), average, GAN-based (GAN), IDMap-MLP, and IDMap-Diff methods in the large-scale evaluation scenario. Divided by the dotted line, the light-blue area on the left shows the results obtained on the MLS dataset only. The dark-blue area on the right presents the addition of the Common Voice dataset on top of the MLS data. The number of utterances and their corresponding speaker counts are presented on the horizontal axis in the format \(\#\text{utterances}/\#\text{speakers}\).}
  \label{fig:large_scale_eer} 
\end{figure*}

\begin{table}[t]
\centering
\caption{Averaged cosine similarities computed across different speaker pairs on the vectors of $\boldsymbol{x}$, $\phi^{\rm 1}$, $\phi^{\rm 2}$, and ${\phi}$.}
\begin{tabular}{c|c|c|c|c}
\Xhline{1pt}
  method & $\boldsymbol{x}$ & $\phi^{\rm 1}$ & $\phi^{\rm 2}$ & ${\phi}$ \\
\hline
IDMap-MLP & 0.1947 & 0.8434 & 0.9907 & 0.9997 \\
\hline
IDMap-Diff & 0.1947 & 0.9415 & 0.9983 & 0.9999 \\
\Xhline{1pt}
\end{tabular}
\label{tab:intra_class_similarity}
\vspace{-0.35cm}
\end{table}

The results show that on the original speaker vector vectors ${\boldsymbol{x}}$, the average speaker similarity across different speaker pairs was 0.1947, indicating high speaker distinction. Speaker similarity increased after passing through the two blocks in the auxiliary processor successively in both theIDMap-MLP and IDMap-Diff models. Finally, the output vector \(\phi\) achieved speaker similarities of 0.9997 and 0.9999 in the two models, respectively, approaching 1, which is the upper bound of cosine similarity. These observations indicate that the auxiliary processor effectively removed speaker-specific information from the original speaker vectors. This enables the use of the speaker vector extracted from any speech utterance as its input during the inference process. Moreover, the cosine similarity of approximately 1 obtained by the output vector \( \phi \) indicates that it lacks speaker-specific information. As such, in the IDMap framework depicted in Fig. \ref{fig:anonymizer}(\subref{fig:anonymizer_overall}), its introduction to the speaker-specific vector $\boldsymbol{u}$ is supposed to provide regularization for generating the vector within the space defined by the speaker vectors.

\subsection{Large-scale anonymization evaluations}
Finally, experiments were conducted in the large-scale anonymization scenario. Both ASV and ASR evaluations were carried out and detailed as follows.

\subsubsection{ASV evaluations}
In the large-scale ASV evaluations, the number of utterances increased from 2,088 to 358,482, originating from 50 and 10,000 speakers, respectively. In each test, the same speakers were used in the enrollment and trial. For each speaker, a maximum of 10 utterances were selected for enrollment, and a maximum of 30 utterances were selected for the trial. The utterances utilized for enrollment and trial were distinct for each speaker. The evaluation trials were configured by pairing every enrollment utterance with every trial utterance across all speakers. Utterances from the same speaker constituted target trials, while those from different speakers constituted nontarget trials. The ASV results are presented in Fig. \ref{fig:large_scale_eer}, measured with EERs. The proposed IDMap-MLP and IDMap-Diff models were compared with the random selection, average, and GAN-based methods. In this comparison, the PSD method was skipped due to its excessive computational cost.

From Fig. \ref{fig:large_scale_eer}, it can be observed that the proposed IDMap framework outperformed the baseline methods in both the IDMap-MLP and IDMap-Diff models across all configurations concerning the number of utterances. This indicates the superiority of the IDMap framework in voice privacy protection in the large-scale scenario. Moreover, regarding the degradation with the increasing number of utterances, the random selection, average, and GAN-based methods exhibited relative 61.8\%, 62.9\%, 40.5\% decreases as the number of utterances increased from 2,088 to 358,482, respectively. In comparison, the relative decrease was 21.1\% in the IDMap-MLP model and 13.7\% in the IDMap-Diff model, much lower than the baseline methods. This comparison demonstrates that the IDMap framework improved the stability of the voice privacy protection capability with an increasing number of generated pseudo-speakers. This suggests that the superiority of the proposed IDMap framework in voice privacy protection was further pronounced in the large-scale scenario. Moreover, between IDMap-MLP and IDMap-Diff, the latter exhibited higher EERs and slower degradation with an increasing number of utterances, indicating enhanced capability and stability in voice privacy protection in the large-scale scenario.

\begin{figure}[t]
  \centering
  \includegraphics[width=1\columnwidth]{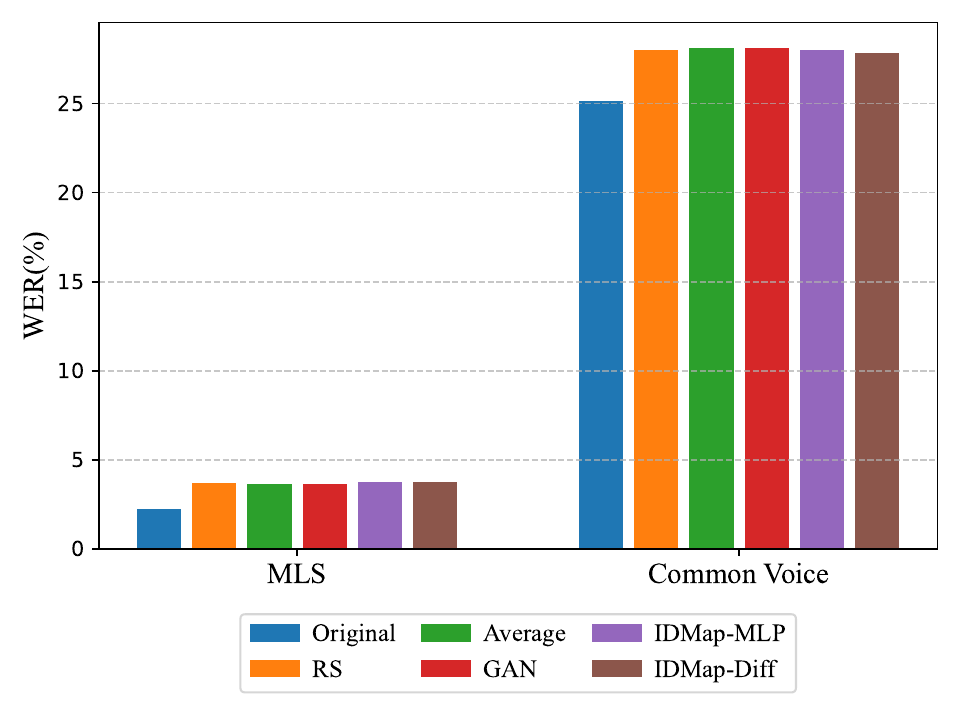}
  \caption{WERs(\%) obtained by the average, GAN-based, IDMap-MLP, and IDMap-Diff methods on MLS and Common Voice, respectively. The results obtained on the original recordings are included.}
  \label{fig:ASR_large}
\end{figure}

\subsubsection{ASR evaluations}

In the ASR evaluations, five utterances were selected from each speaker for evaluation. The evaluations were performed on MLS and Common Voice separately. The WERs obtained from the random selection, average, GAN-based, IDMap-MLP, and IDMap-Diff methods are presented in Fig. \ref{fig:ASR_large}, alongside those obtained from the original recordings. From the results, both the proposed IDMap-MLP and IDMap-Diff models obtained no higher WERs than the baseline methods on both the MLS and Common Voice datasets. These results further demonstrate the efficacy of the proposed IDMap framework in linguistic content preservation.

\subsubsection{Capacity}

\begin{figure}[t]
  \centering
    \includegraphics[width=0.95\columnwidth]{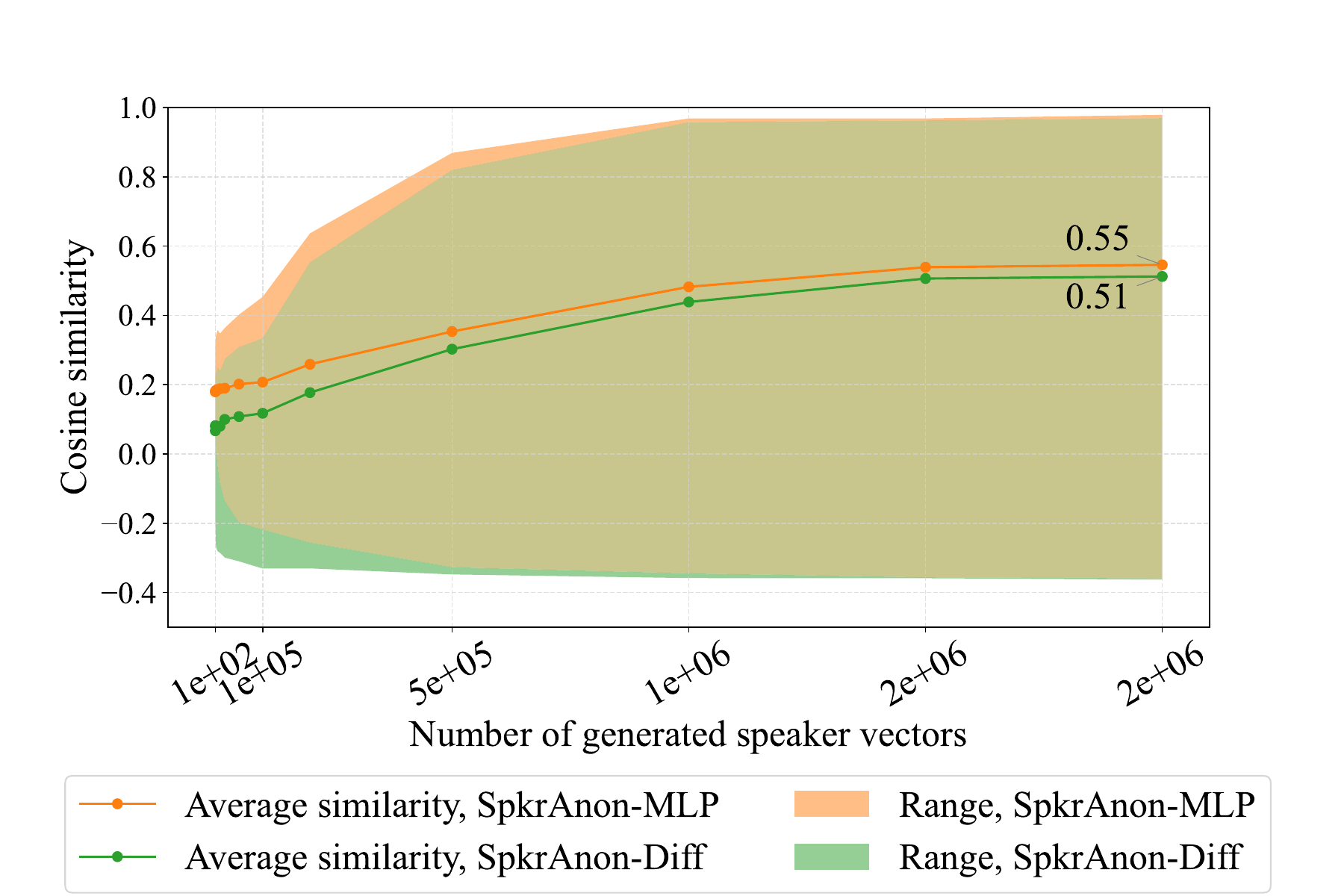}
  \caption{The average cosine similarity between speaker vectors generated by IDMap-MLP and IDMap-Diff, respectively. The number of generated speaker vectors increases from 100 to 2,000,000. The range between the minimum and maximum values is provided for each.} 
  \label{fig:cos_pseudo}
\end{figure} 

Lastly, the capacity of the IDMap-MLP and IDMap-Diff models was examined considering the uniqueness of the generated pseudo-speaker vectors. To this end, a number of speaker vectors were generated with the IDMap-MLP and IDMap-Diff models, respectively, ranging from 100 to 2,000,000. Given a speaker vector, pairs were formed between it and all the other speaker vectors. Cosine similarity was calculated within each pair. Higher similarity among the generated speaker vectors indicated lower distinctiveness, which, when applied to anonymized speech generation, may lead to lower uniqueness among the pseudo-speakers. The averages of the cosine similarity values obtained from the speaker vectors generated by the IDMap-MLP and IDMap-Diff models are presented in Fig. \ref{fig:cos_pseudo}. Besides, the ranges, representing the interval between the lowest and highest values, are shown for the two models respectively. As shown in the figure, an increase in the number of generated speaker vectors led to an increase in cosine similarities, indicating a decrease in uniqueness IDMap for both models. Between the IDMap-MLP and IDMap-Diff models, the latter demonstrated a higher level of uniqueness, further justifying its superiority in speaker vector generation. Moreover, with the increase in the generated speaker vectors, the average cosine similarity values saturated at 0.55 and 0.51 for IDMap-MLP and IDMap-Diff, respectively. The cosine similarities calculated from the speaker vectors generated by the IDMap-MLP model ranged from -0.35 to 0.98, while those from the IDMap-Diff model ranged from -0.37 to 0.97.

\section{Conclusions}
\label{sec.5}
This paper introduces the IDMap framework for pseudo-speaker vector generation, which establishes a mapping from speaker identity index to speaker vector with a feedforward architecture. It enables the generation of a speaker vector given a speaker identity index. In the anonymization process, a new pseudo-speaker is generated by assigning a speaker identity index that is distinct from those previously utilized, thereby achieving uniqueness in the pseudo-speaker. Concerning the generator module within the framework, this study examines two specifications of the framework: IDMap-MLP and IDMap-Diff. The proposed models were examined in both small- and large-scale scenarios based on the number of generated pseudo-speakers. The small-scale experimental evaluations conducted on the LibriSpeech dataset demonstrate the effectiveness of the proposed IDMap framework in enhancing the pseudo-speaker uniqueness, thereby improving the voice privacy protection capability, while at a reduced computation cost. The superiority of the IDMap framework was further validated in large-scale evaluations on the MLS and Common Voice datasets, demonstrating enhanced stability in voice privacy protection capability as the number of generated pseudo-speakers increased.

\bibliographystyle{IEEEtran}
\bibliography{mybib}

@inproceedings{casanova2022yourtts,
  title={{YourTTS}: {Towards} zero-shot multi-speaker {TTS} and zero-shot voice conversion for everyone},
  author={Casanova, Edresson and Weber, Julian and Shulby, Christopher D and Junior, Arnaldo Candido and G{\"o}lge, Eren and Ponti, Moacir A},
  booktitle={Proc. ICML},
  pages={2709--2720},
  year={2022},
}

@article{ning2019review,
  title={A review of deep learning based speech synthesis},
  author={Ning, Yishuang and He, Sheng and Wu, Zhiyong and Xing, Chunxiao and Zhang, Liang-Jie},
  journal={Applied Sciences},
  volume={9},
  number={19},
  pages={4050},
  year={2019},
  publisher={MDPI}
}

@inproceedings{se-resnet,
  title={Squeeze-and-excitation networks},
  author={Hu, Jie and Shen, Li and Sun, Gang},
  booktitle={Proc. CVPR},
  pages={7132--7141},
  year={2018}
}

@article{turner2022generating,
  title={Generating identities with mixture models for speaker anonymization},
  author={Turner, H. and Lovisotto, G. and Martinovic, I.},
  journal={Computer Speech \& Language},
  volume={72},
  pages={101318},
  year={2022}
}

@inproceedings{ECAPATDNN,
title={{ECAPA-TDNN}: Emphasized Channel Attention, Propagation and Aggregation
in {TDNN} Based Speaker Verification},
author={Desplanques, B. and Thienpondt, J. and Demuynck, K.},
booktitle={Proc. InterSpeech},
year={2020},
pages={3830--3834}
}

@ARTICLE{voice_conversion_overview,
  author={Sisman, B. and Yamagishi, J. and King, S. and Li, H.},
  journal={IEEE/ACM Transactions on Audio, Speech, and Language Processing}, 
  title={An overview of voice conversion and its challenges: from statistical modeling to deep learning}, 
  year={2021},
  volume={29},
  number={},
  pages={132-157},
  publisher={IEEE}
}

@inproceedings{snyder17_interspeech,
  author={D. Snyder and Daniel Garcia-Romero and Daniel Povey and Sanjeev Khudanpur},
  title={Deep neural network embeddings for text-independent speaker verification},
  year=2017,
  booktitle={Proc. Interspeech},
  pages={999--1003}
}

@article{li2022recent,
  title={Recent advances in end-to-end automatic speech recognition},
  author={Li, Jinyu and others},
  journal={APSIPA Transactions on Signal and Information Processing},
  volume={11},
  number={1},
  year={2022},
  publisher={Now Publishers, Inc.}
}

@article{kong2020hifi,
  title={{HiFi-GAN}: {Generative} adversarial networks for efficient and high fidelity speech synthesis},
  author={Kong, Jungil and Kim, Jaehyeon and Bae, Jaekyoung},
  journal={Advances in neural information processing systems},
  volume={33},
  pages={17022--17033},
  year={2020}
}

@inproceedings{wang2018style,
  title={{Style tokens: Unsupervised style modeling, control and transfer in end-to-end speech synthesis}},
  author={Wang, Yuxuan and Stanton, Daisy and Zhang, Yu and Ryan, RJ-Skerry and others},
  booktitle={Proc. ICML},
  pages={5180--5189},
  year={2018},
  organization={PMLR}
}

@inproceedings{sun2016phonetic,
  title={Phonetic posteriorgrams for many-to-one voice conversion without parallel data training},
  author={Sun, Lifa and Li, Kun and Wang, Hao and Kang, Shiyin and Meng, Helen},
  booktitle={Proc. ICME},
  pages={1--6},
  year={2016},
  organization={IEEE}
}

@inproceedings{panayotov2015librispeech,
  title={Librispeech: {An} {ASR} corpus based on public domain audio books},
  author={Panayotov, Vassil and Chen, Guoguo and Povey, Daniel and Khudanpur, Sanjeev},
  booktitle={Proc. ICASSP},
  pages={5206--5210},
  year={2015},
  organization={IEEE}
}

@inproceedings{libritts,
title	= {{LibriTTS}: {A} Corpus Derived from Librispeech for Text-to-Speech},
author	= {H. Zen and R. Clark and R. J. Weiss and V. Dang and Y. Jia and Y. Wu and Y. Zhang and Z. Chen},
year	= {2019},
pages={1526–1530},
booktitle	= {Proc. Interspeech}
}

@article{IEMOCAP,
  title={{IEMOCAP}: {Interactive} emotional dyadic motion capture database},
  author={Busso, C. and Bulut, M. and Lee, C.C. and Kazemzadeh, A. and Mower, E. and Kim, S. and Chang, J.N. and Lee, S. and Narayanan, S.S.},
  journal={Journal of Language Resources and Evaluation},
  volume={42},
  number={4},
  pages={335--359},
  year={2008}
}

@inproceedings{esd,
  title={Seen and unseen emotional style transfer for voice conversion with a new emotional speech dataset},
  author={Zhou, Kun and Sisman, Berrak and Liu, Rui and Li, Haizhou},
  booktitle={Proc. ICASSP},
  pages={920--924},
  year={2021},
  organization={IEEE}
}

@article{vpc2024,
  title={The {VoicePrivacy 2024 Challenge} Evaluation Plan},
  author={Tomashenko, Natalia and Miao, Xiaoxiao and Champion, Pierre and Meyer, Sarina and Wang, Xin and Vincent, Emmanuel and Panariello, Michele and Evans, Nicholas and Yamagishi, Junichi and Todisco, Massimiliano},
  journal={arXiv preprint arXiv:2404.02677},
  year={2024}
}

@inproceedings{xvector-based,
  author = {Fang, F. and Wang, X. and Yamagishi, J. and Echizen, I. and Todisco, M. and Evans, N. and Bonastre, J.-F.},
  title = {{Speaker} Anonymization Using X-vector and Neural Waveform Models},
  booktitle = {Proc. ISCA Workshop on Speech Synthesis },
  year = {2019},
  pages = {155-160},
  doi = {10.21437/SSW.2019-28}
}

@inproceedings{vq-bn,
   title={Are disentangled representations all you need to build speaker anonymization systems?},
   booktitle={Proc. Interspeech},
   author={Champion, Pierre and Jouvet, Denis and Larcher Anthony},
   year={2022},
    pages={2793--2797}
 }

@article{ohnn,
  title={Speaker anonymization using orthogonal {Householder} neural network},
  author={Miao, Xiaoxiao and Wang, Xin and Cooper, Erica and Yamagishi, Junichi and Tomashenko, Natalia},
  journal={IEEE/ACM Transactions on Audio, Speech, and Language Processing},
  year={2023},
}

@inproceedings{vq-gst,
  title={{A} Voice Anonymization Method Based on Content and Non-content Disentanglement for Emotion Preservation},
  author={Gu, Wenju and Liu, Zeyan and Chen, Liping and Wang, Rui and Guo, Chenyang and Guo, Wu and Lee, Kong Aik and Ling, Zhen-Hua},
  booktitle={Proc. SPSC},
  pages={116--120},
  year={2024}
}

@inproceedings{mmer,
  author={Sreyan Ghosh and Utkarsh Tyagi and S Ramaneswaran and Harshvardhan Srivastava and Dinesh Manocha},
  title={{MMER: Multimodal multi-task learning for speech emotion recognition}},
  year=2023,
  booktitle={Proc. Interspeech},
  pages={1725--1729},
}

@inproceedings{nac,
  title={{Speaker} anonymization using neural audio codec language models},
  author={Panariello, Michele and Nespoli, Francesco and Todisco, Massimiliano and Evans, Nicholas},
  booktitle={Proc. ICASSP},
  pages={4725--4729},
  year={2024},
}

@ARTICLE{psd,
  author={Chen, Liping and Gu, Wenju and Lee, Kong Aik and Guo, Wu and Ling, Zhen-Hua},
  journal={IEEE Transactions on Audio, Speech and Language Processing}, 
  title={Pseudo-Speaker Distribution Learning in Voice Anonymization}, 
  year={2025},
  volume={33},
  pages={272--285}
}

@article{svd,
  title={{Distinctive} and Natural Speaker Anonymization via Singular Value Transformation-Assisted Matrix},
  author={Yao, Jixun and Wang, Qing and Guo, Pengcheng and Ning, Ziqian and Xie, Lei},
  journal={IEEE/ACM Transactions on Audio, Speech, and Language Processing},
  volume={32},
  pages={2944--2956},
  year={2024}
}

@inproceedings{vox1,
 title={{VoxCeleb: {A} large-scale speaker identification dataset}},
 author={Nagrani, A. and Chung, J. S. and Zisserman, A.},
 booktitle={Proc. Interspeech},
 pages={2616--2620},
 year={2017}
}

@inproceedings{vox2,
 title={{VoxCeleb2: Deep speaker recognition}},
 author={Chung, J. S. and Nagrani, A. and Zisserman, A.},
 booktitle={Proc. Interspeech},
 pages={1086--1090},
 year={2018}
}

@inproceedings{average,
 title={{Design choices for x-vector based speaker anonymization}},
 author={Srivastava, Brij Mohan Lal and Tomashenko, Natalia and Wang, Xin and Vincent, Emmanuel and Yamagishi, Junichi and Maouche, Mohamed and Bellet, Aur{\'e}lien and Tommasi, Marc},
 booktitle={Proc. Interspeech},
 pages={1713--1717},
 year={2020}
}

@article{cohort-speaker,
  title={{Privacy and utility of x-vector based speaker anonymization}},
  author={Srivastava, Brij Mohan Lal and Maouche, Mohamed and Sahidullah, Md and Vincent, Emmanuel and Bellet, Aur{\'e}lien and Tommasi, Marc and Tomashenko, Natalia and Wang, Xin and Yamagishi, Junichi},
  journal={IEEE/ACM Transactions on Audio, Speech, and Language Processing},
  volume={30},
  pages={2383--2395},
  year={2022}
}

@inproceedings{GAN,
  title={Anonymizing Speech with Generative Adversarial Networks to Preserve Speaker Privacy},
  author={Meyer, S. and Tilli, P. and Denisov, P. and Lux, F. and Koch, J. and Vu, N. T.},
  booktitle={Proc. SLT},
  pages={912--919},
  year={2023}
}

@inproceedings{noe20_interspeech,
  author={Paul-Gauthier Noé and Jean-François Bonastre and Driss Matrouf and N. Tomashenko and Andreas Nautsch and Nicholas Evans},
  title={Speech Pseudonymisation Assessment Using Voice Similarity Matrices},
  year=2020,
  booktitle={Proc. Interspeech},
  pages={1718--1722}
}

@inproceedings{wgan-qc,
  title={Wasserstein {GAN} with quadratic transport cost},
  author={Liu, Huidong and Gu, Xianfeng and Samaras, Dimitris},
  booktitle={Proc. ICCV},
  pages={4832--4841},
  year={2019}
}

@inproceedings{commonvoice,
  author = {Ardila, R. and Branson, M. and Davis, K. and Henretty, M. and Kohler, M. and Meyer, J. and Morais, R. and Saunders, L. and Tyers, F. M. and Weber, G.},
  title = {{Common Voice}: A Massively-Multilingual Speech Corpus},
  booktitle = {Proceedings of the 12th Conference on Language Resources and Evaluation (LREC 2020)},
  pages = {4211--4215},
  year = 2020
}

@inproceedings{mls,
  title     = {{MLS}: A Large-Scale Multilingual Dataset for Speech Research},
  author    = {Vineel Pratap and Qiantong Xu and Anuroop Sriram and Gabriel Synnaeve and Ronan Collobert},
  booktitle = {Proc. Interspeech},
  pages     = {2757--2761},
  year      = {2020},
  doi       = {10.21437/Interspeech.2020-2826},
}

@inproceedings{DiffVC,
  title     = {Diffusion-Based Voice Conversion with Fast Maximum Likelihood Sampling Scheme},
  author    = {Vadim Popov and Ivan Vovk and Vladimir Gogoryan and Tasnima Sadekova and Mikhail Kudinov and Jiansheng Wei},
  booktitle = {Proc. ICLR},
  year      = {2022}
}

@article{pcg,
  title={{PCG}: A family of simple fast space-efficient statistically good algorithms for random number generation},
  author={O’neill, Melissa E},
  journal={ACM Transactions on Mathematical Software},
  volume={204},
  year={2014}
}

@INPROCEEDINGS{librilight,
  author={J. {Kahn} and M. {Rivière} and W. {Zheng} and E. {Kharitonov} and Q. {Xu} and P. E. {Mazaré} and J. {Karadayi} and V. {Liptchinsky} and R. {Collobert} and C. {Fuegen} and T. {Likhomanenko} and G. {Synnaeve} and A. {Joulin} and A. {Mohamed} and E. {Dupoux}},
  booktitle={Proc. ICASSP}, 
  title={Libri-Light: A Benchmark for ASR with Limited or No Supervision}, 
  year={2020},
  pages={7669-7673}
}

@inproceedings{switchboard,
  title={SWITCHBOARD: Telephone speech corpus for research and development},
  author={Godfrey, John J and Holliman, Edward C and McDaniel, Jane},
  booktitle={Acoustics, speech, and signal processing, ieee international conference on},
  volume={1},
  pages={517--520},
  year={1992},
  organization={IEEE Computer Society}
}

@inproceedings{fisher,
  title={The Fisher corpus: A resource for the next generations of speech-to-text.},
  author={Cieri, Christopher and Miller, David and Walker, Kevin},
  booktitle={LREC},
  volume={4},
  pages={69--71},
  year={2004}
}

@article{xivector,
  title={{Xi-Vector} embedding for speaker recognition},
  author={Lee, Kong Aik and Wang, Qiongqiong and Koshinaka, Takafumi},
  journal={IEEE Signal Processing Letters},
  volume={28},
  pages={1385--1389},
  year={2021},
  publisher={IEEE}
}

@article{numpy,
  title={Array programming with NumPy},
  author={Harris, Charles R and Millman, K Jarrod and Van Der Walt, St{\'e}fan J and Gommers, Ralf and Virtanen, Pauli and Cournapeau, David and Wieser, Eric and Taylor, Julian and Berg, Sebastian and Smith, Nathaniel J and others},
  journal={Nature},
  volume={585},
  number={7825},
  pages={357--362},
  year={2020},
}


\end{document}